\documentclass[aps,pra,twocolumn,floatfix,showpacs]{revtex4-1}
\usepackage{graphicx,amsfonts}
\usepackage{amsmath,amscd,amsfonts,amssymb,color}
\newcommand{\bra}[1]{\left\langle{#1}\right\vert}
\newcommand{\ket}[1]{\left\vert{#1}\right\rangle}
\begin{document}
\title{Estimation of quantum states by weak and projective
measurements}
\author{Debmalya Das}
\email{debmalya@iisermohali.ac.in}
\affiliation{Department of Physical Sciences, Indian Institute of
Science Education
\& Research (IISER) Mohali, Sector-81, SAS Nagar, Manauli P.O.
140306, Punjab, India.}
\author{Arvind}
\email{arvind@iisermohali.ac.in}
\affiliation{Department of Physical Sciences, Indian Institute of
Science Education
\& Research (IISER) Mohali, Sector-81, SAS Nagar, Manauli P.O.
140306, Punjab, India.}
\begin{abstract}
We explore the possibility of using ``weak'' measurements to
carry out quantum state tomography via numerical simulations.
Given a fixed number of copies of identically
prepared states of a qubit, we perform state tomography
using weak as well as projective measurements.  Due to the
collapse of the state after measurement, we cannot reuse
the state after a projective measurement.  If the coupling
strength between the quantum system and the measurement
device is made weaker, the disturbance caused to the state
can be lowered.  This then allows us to reuse the same
member of the ensemble for further measurements and thus
extract more information from the system. However, this
happens at the cost of getting imprecise information from
the first measurement. We implement this scheme for a single
qubit and show that under certain circumstances, it can
outperform the projective measurement-based tomography
scheme. This opens up the possibility of new ways of
extracting information from quantum ensembles. We study the
efficacy of this scheme for different coupling strengths,
and different ensemble sizes.
\end{abstract} 
\maketitle
\section{Introduction} \label{introduction}
Measurement in quantum physics has a very different
connotation as compared to that in classical physics.
Measurement invariably disturbs the quantum system, and we
say that information comes at a certain cost.  The most
commonly encountered quantum measurements are projective
measurements, wherein the state collapses into one of the
eigenvectors of the observable being measured. There is no
further information that one can obtain by making a repeated
measurement on this collapsed state. Alternatively, one
could conceive of ``weak'' or ``unsharp'' measurements,
where the coupling of the apparatus with the system is weak,
and only a limited amount of noise is introduced.
Consequently, the information obtained from this 
measurement is also limited.  However, there is a
possibility of recycling the state and making further
measurements on it, which may reveal more information about
the state.

Ideal state estimation would require an infinite number of
copies of identically prepared states, however, in reality
we always have a finite ensemble. Therefore, it would be
interesting to explore the possibilities of reducing the
size of the ensemble required to achieve a certain amount of
fidelity of state estimation.  We explore the possibility of
carrying out state tomography on finite ensembles using weak
measurements, where we recycle the state to extract
information about more than one observable.
An unsharp or weak measurement is achieved when the
apparatus system coupling is weak compared to the initial
spread of the pointer state wave functions. This can be
achieved by reducing the coupling strength or by preparing
the initial pointer states in sufficiently wide
wave functions.  
For such weak measurements the state of the
system does not collapse fully, and the state can still be
used to extract more information. Such schemes involving
weak or unsharp or fuzzy measurements have been proposed in
the literature~\cite{Busch,kraus1983,AliFuzzy,Prugovecki1,Prugovecki2,
Diosi}.
Weak measurement has an interesting property
that although it yields very little information~\cite{brun},
the state is correspondingly disturbed very little. 
However, in such a measurement,
the pointer positions corresponding to different eigenvalues
of the observer being measured could overlap, leading to an
ambiguity. A certain region of the pointer position may
therefore have to be discarded in order to reduce the
ambiguity in the measurement. Thus we have two parameters,
namely the strength of the measurement and the discard
parameter, over which we can optimize the performance of the
measurement scheme.  This on the one hand provides a novel
way of extracting information from a quantum system, and on
the other hand, may lead to improvement in fidelity over
projective measurements. 
It may be noted that it is the interplay between the initial
state of the pointer and the coupling strength which defines
a weak (unsharp) measurement. In fact if we are able to
prepare a very narrow initial state of the pointer, even a
weak coupling strength can lead to a projective measurement.
Weak measurements are also associated with ``weak values'',
which require the notion of post-selection~\cite{AAV,Sudweak,
Toll,Joz}. This process of post-selection leads to throwing
away data and can lead to suboptimal use of information from
a measurement~\cite{Combes,Ferrie}. In
our work we use weak or unsharp measurements without
postselection.
Although all quantum measurements (projective,
non-projective, weak etc) can be seen as Positive Operator
Valued Measures (POVM), it is important to know the 
details of a measurement scheme. A POVM can also be
interpreted as a projective measurement on a larger Hilbert
space~\cite{NC,P3,brun}.  For a finite ensemble the upper
bound on the amount of information extractable is 
available~\cite{MassPop1}. The cost of  information extraction 
from quantum systems in terms of disturbance caused has
also been explored in the context of weak measurements
~\cite{Ueda,Branciard,Cheong}.

A good way to represent pure as well as mixed 
states of a single qubit is to
use the Bloch sphere~\cite{NC,sakurai}. 
The Bloch sphere is a unit ball and every
point on and inside the sphere represents a quantum state of
the qubit. The state corresponding to the point 
$(x,y,z)$ is given by
\begin{equation}
 \rho=\frac{1}{2}\left(I+\vec{n}.\vec{\sigma}\right)
\end{equation}
where $\hat{n}=x\hat{x}+y\hat{y}+z\hat{z}$  is a  vector with 
$x=\langle \sigma_x\rangle$, $y=\langle \sigma_y\rangle$ and 
$z=\langle \sigma_z\rangle$. The pure states correspond to
the case when the point lies on the surface and in that case
$\vec{n}$ is a unit vector. 
The expectation values of $\sigma_x$, $\sigma_y$ and
$\sigma_z$ serve as a direct means to calculate the values of
$(x,y,z)$. Therefore, to carry out state estimation of a
given state of a
single qubit, we need to estimate
the numbers $(x,y,z)$.

The efficacy of any state tomography procedure is determined
by the closeness of the estimated state to the state being
tomographed. This requires an appropriate fidelity measure.
Since we are dealing with general states of a qubit we
consider the distance between the estimated state and the
original state on the Bloch sphere as a measure of the fidelity
of the tomography scheme. 
Let us assume that the estimated values of 
$(x,y,z)$ are $(x_{est},y_{est},z_{est})$ for a given
estimation scheme.  
We define a measure of fidelity as 
\begin{equation}
\label{fidelity}
 f=1-\left[\left(x-x_{est}\right)^2 +
\left(y-y_{est}\right)^2 + \left(z-z_{est}\right)^2\right]
\end{equation}
The fidelity is a measure of the distance between the
original state and the estimated state. For a perfect estimation $f$
is equal to one. The amount by which $f$ is less than one
measures the departure of the estimate from the original
state. We will use this measure throughout this
paper, to measure the efficacy of the state estimation schemes.

This work explores state reconstruction for pure and mixed
states of a qubit using weak measurements and compares the
efficacy of this scheme with that using projective
measurements.  Since state tomography requires an ensemble
of identically prepared states, we have assumed finite
ensembles and calculated the dependence of the fidelity of
the tomography scheme as a function of ensemble size in both
cases. We show that under certain circumstances, weak
measurements with state recycling can be a better tool for
state reconstruction.  This we believe, extends the scope of
extracting information from quantum systems at a reduced
cost. 

The material in this paper is arranged as follows: In
section~\ref{weak-measurement} we describe weak
measurements. Section~\ref{procedure} details the tomography
procedure using weak measurements on finite ensembles. In
this section the main results of the simulation are
presented.
Section~\ref{conc} contains some discussion and concluding
remarks.
\section{Weak Measurements in Quantum Mechanics}
\label{weak-measurement}
The process of gaining information from a quantum system
typically requires an apparatus with distinct classical
(macroscopic) pointer positions to interact with the quantum
system followed by a read out of the pointer positions.  A
useful model of this process is available due to von
Neumann. Although originally this model was constructed for
strong (projective) measurements~\cite{johnV} it has wider applications
and can also be applied to weak
measurements~\cite{Busch,kraus1983,AliFuzzy,Prugovecki1,Prugovecki2,
Diosi,AAV,Sudweak}.

Consider the measurement of an observable $A$ of a quantum
system with eigenvectors $\{\vert a_j\rangle\}$ and
eigenvalues $\{a_j\}$, $j=1\cdots n$. Imagine an apparatus
with continuous pointer positions described by a variable
$q$ and its conjugate variable $p$ such that $[q,p]=i$.
The initial state of the measuring device has an initial
spread of $\Delta q$ with its initial Gaussian quantum state
$\ket{\phi_{in}}$ centered around zero given by 
\begin{equation}
\ket{\phi_{in}}=\left(\frac{\kappa}{2\pi}\right)^{\frac{1}{4}}\int_{-\infty}
^\infty dq \,e^{-\frac{\kappa q^2}{4}}\ket{q}
\label{pointer_in}
\end{equation}
where $\kappa=\frac{1}{\left(\Delta q\right)^2}$ and we
have taken $\hbar=1$.  The system and the measuring device
are made to interact by means of a Hamiltonian,
\begin{equation}
\label{hamiltonian}
 H=g\delta\left(t-t^\prime\right)A\otimes p
\end{equation}
where $p$ is the momentum conjugate to the variable $q$,
and $g$ is the coupling strength. The
Hamiltonian is so chosen that the system and the device get
a kick and interact momentarily at $t=t^\prime$. Let the
initial state $\ket{\psi_{in}}$ of the system be written in
terms of the eigenstates $\ket{a_1},
\ket{a_2},.....,\ket{a_n}$ of the operator $A$.
\begin{equation}
 \ket{\psi_{in}}=\sum_{i=1}^n c_i \ket{a_i}
\end{equation}
The joint evolution of the system and the measuring device
under the coupling Hamiltonian gives an entangled
state for $t> t^{\prime}$
\begin{eqnarray}
&&e^{-i \int H
dt}\ket{\psi_{in}}\otimes\ket{\phi_{in}}=\nonumber \\
&&\left(\frac{\kappa }{2\pi}\right)^{\frac{1}{4}}\sum_{i=1}^n
\int_{-\infty}^\infty dq c_i \,e^{-\frac{\kappa \left(q-ga_i\right)^2}{4}}\ket{a_i}\otimes\ket{q}
\label{grand}
\end{eqnarray}
The above state consists of a series of Gaussians centered
at $g a_1, g a_2,\cdots,g a_n$ for the pointer entangled
with corresponding eigenstates $\vert a_1\rangle,\vert
a_2\rangle \cdots \vert a_n\rangle$ of the system.  At this
stage we invoke the ``classicality'' of the
apparatus, because of
the fact that only one of the pointer positions actually
shows up. This requires the collapse of the wave function
which is brought in as something natural for the classical
apparatus!  Thus the  process is completed with the meter
showing only one of the $g a_i$s. Consequently, the system
state collapses into the corresponding eigenstate $\vert
a_i\rangle$.  The above analysis holds good only if the
Gaussians are well separated or distinct. In case they
overlap, which can happen if the coupling strength $g$ is
small or the initial spread in the pointer state given by
$1/\kappa$  is large, the scenario changes~\cite{Busch,AAV,Marc}.
This is called the weak or unsharp measurement regime.  
Weak measurements
have been employed in developing recipes for the violation
of Bell inequalities~\cite{Marc} and Leggett Garg
inequalities~\cite{LGineq}.  These have also been recently
used to study super quantum discord~\cite{patisingh1,
patisingh2}. 

In the treatment of weak measurement given by Aharonov,
Albert and Vaidmann (AAV), a subsequent projective
measurement of a second observable is carried out, followed
by a post-selection of the output state into one of the
eigenstates of the second observable.  
However, we take a different approach in our work, where we
do not do any post-selection i.e. we consider {\em weak
measurements without weak values}.

How exactly do we carry out the weak measurement?  How much
is the effect of a weak measurement on the system?  If we
carry out weak measurements on all the members of an
identically prepared ensemble, what happens to such an
ensemble? We illustrate these points by taking an example.
Consider a measurement of $\sigma_z$ ($z$ component of spin)
of a qubit in a fixed quantum state. Following the general
prescription given in Equation~(\ref{hamiltonian}) we write
the interaction Hamiltonian 

\begin{equation}
\label{interaction}
 H=g\delta\left(t-t^\prime\right)\sigma_z\otimes p
\end{equation}
assuming the initial state of the pointer to be
the same as that
given in Equation~(\ref{pointer_in}).  The qubit
is taken to be 
in a pure
state given by 
\begin{equation}
\vert \psi_{in} \rangle = \cos{\frac{\alpha}{2}} \vert 0
\rangle+ \sin{\frac{\alpha}{2}}\vert 1 \rangle
\end{equation}
where $\vert 0 \rangle$ and $\vert 1 \rangle$ are the eigenstates of $\sigma_z$ with eigen values $+1$ and $-1$
respectively.
The combined state of
the system and the pointer after the interaction is given by 
taking a special case of Equation~(\ref{grand}) 
\begin{eqnarray}
\vert \psi_{out}\rangle =&&
\left(\frac{\kappa }{2\pi}\right)^{\frac{1}{4}}
\int_{-\infty}^\infty dq \cos{\frac{\alpha}{2}} 
\,e^{-\frac{\kappa
\left(q-g\right)^2}{4}}\ket{0}\otimes\ket{q}\nonumber \\
&&+\left(\frac{\kappa }{2\pi}\right)^{\frac{1}{4}}
\int_{-\infty}^\infty dq \sin{\frac{\alpha}{2}} 
\,e^{-\frac{\kappa 
\left(q+g\right)^2}{4}}\ket{1}\otimes\ket{q}\nonumber \\
\label{grand_1}
\end{eqnarray}
At this stage the apparatus and the system are in an
entangled state.
An observation of the apparatus will lead to values
whose distribution is determined by the above state. It is
clear from Equation~(\ref{grand_1}) that the distribution of
values of the apparatus is a Gaussian centered around $+g$
for the system input state $\vert 0 \rangle$ and is a
Gaussian centered around $-g$ 
for the system input state $\vert 1 \rangle$.
The width of the Gaussian in  each case is given by
$1/{\kappa}$. By tuning the parameter $\epsilon=\kappa g $
we can change the nature of the measurement in terms of its
strength. In our
work we have taken $g=1$ so that we have $\epsilon=\kappa$.
For large
values of $\epsilon$ we have a projective measurement, where
the pointer distributions are well separated for the states 
$\vert 0 
\rangle$ and $\vert 1 \rangle$. Therefore, each reading of
the pointer tells us exactly what the state of the system is
after the measurement. By repeatedly measuring the same
observable we can calculate the expectation value of the
observable. The state collapses completely in each
measurement
and there is no question of re-using these states.
However, when the value of $\epsilon$ is small we have two
Gaussians that overlap. From an observation of
the pointer we do not learn with certainty as to what value
to assign to the system spin $z$ component. The pointer
positions are weakly correlated with the eigenstates of
$\sigma_z$. The state is only partially affected and there
is a possibility of re-using the state.  
The effect of the weak measurement in this
case can be explicitly calculated and it turns out that there is
very little change in the state of the system.
The final state of the system can be calculated
by taking the state in Equation~(\ref{grand}) and then
taking a partial trace over
the apparatus's degrees of freedom giving us the final  mixed state
corresponding to the system alone:
\begin{equation}
\rho_f=\frac{1}{2}\begin{pmatrix}
  1+\cos{\alpha}&&
(1-\frac{\epsilon}{8})\sin{\alpha}\\
  (1-\frac{\epsilon}{8})\sin{\alpha}&&
1-\cos{\alpha}
                    \end{pmatrix}
\end{equation}
Since $\epsilon$ is small we can conclude that the
disturbance caused to the system is also small.
Furthermore,
the disturbance can be controlled by changing $\epsilon$.

A recent work by Rozema et. al. suggests some new
possibilities that weak measurements can offer with respect to
Heisenberg's uncertainty relation and the disturbance caused
to the state~\cite{distwm}.  Oreshkov et. al., in 2005,
wrote down a weak measurement POVM and showed that any
generalized measurement can be decomposed into a sequence of
weak measurements, without using an ancilla~\cite{wmPOVM}.
Lundeen et. al. recently came up with a method employing
weak values to
directly measure the wave function of a quantum system in a
pure state~\cite{Qwavefn} and followed it up with a method
to measure any general state~\cite{GenQwavefn}.
For some further developments in this
regard see~\cite{phyQwavefn}. Unsharp measurements have also
been used to make sequential measurements on a single
qubit~\cite{Diosi}. Other examples of quantum
state tomography with weak measurements can be found in
~\cite{sttomowm,sswm,sswm1}. An approach to perform quantum state
tomography using weak measurement POVMs was introduced by
Hofmann~\cite{Hofmann}. 
\section{Quantum State Estimation of a single qubit}
\label{procedure}
We now turn to the question of using weak measurements with
state recycling for the problem of state estimation of a
single qubit. 

\subsection{The scheme}
\begin{figure}[h]
\includegraphics{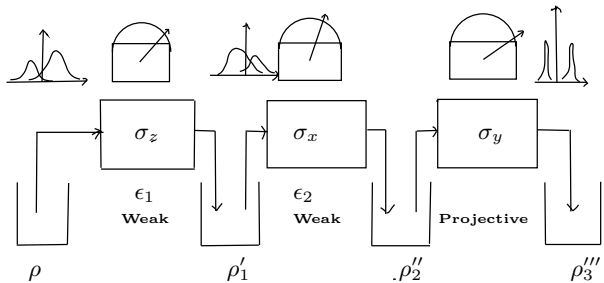}
\caption{\label{flowchart}
The schematic diagram of our scheme where we pick a copy of
the qubit in state $\rho$ from the first box and perform the
measurement of $\sigma_z$ weakly defined through the system
apparatus coupling strength $\epsilon_1$. The state after
this measurement changes to $\rho^{\prime}_1$, on which we
perform the measurement of $\sigma_x$ again weakly defined
through the coupling strength $\epsilon_2$. The state now
changes to $\rho^{\prime \prime}_2$ on which we perform a
projective measurement of $\sigma_y$. The state after the
projective measurement is $\rho^{\prime\prime\prime}_3$ and
we discard this copy since no information can be extracted
from the ensemble.  The overlapping Gaussians in the first
two cases indicate that the measurement is weak while the
non-overlapping outcomes in the last case indicate that the
measurement is projective in nature.  }
\end{figure}
In our prescription, we consider a finite
size ensemble of pure
or mixed states of a qubit.
On every member of the ensemble we
carry out a $\sigma_z$ measurement whose strength is defined
by the parameter $\epsilon_1$.  We record the meter reading
in each case and keep the modified states after measurements to
obtain a changed ensemble. This new ensemble is now used to
measure $\sigma_x$ in the same way but with a coupling
strength $\epsilon_2$. Finally the resultant ensemble is
used to carry out  projective measurement of $\sigma_y$ on
its members. The first two measurements are weak while the
last measurement is strong or projective. 
To avoid statistical errors the results are  averaged 
over many runs. 
The entire process is summarized in
Figure~\ref{flowchart}.
For both the weak measurements,
consider a regime in which $\epsilon$ is neither too large
to make the measurement projective, nor too small, as is
done in traditional weak measurements. For such values of
$\epsilon$, the two Gaussians, representing the pointer value
distributions for the two eigen values of the observable,
overlap partially with each
other. When there is no overlap, a meter reading
unambiguously indicates an outcome and we have a projective
measurement.  A meter reading corresponding to a point in the
overlap region cannot be reliably correlated with the system
being in one or the other eigenstate.
To reduce this difficulty, let us define a
region, midway between the centers of the two Gaussians, of
width $2a$. We call it the discard region, which means that
any pointer reading which falls in this region is rejected.
For the case where we measure $\sigma_z$, 
all readings where the pointer position is to the right of this
region are interpreted as indicating the value of $\sigma_z$
to be  $+1$ while the ones on the left of this region are
interpreted as $-1$. Even when the outcome is discarded, the
member of the ensemble is not rejected, but is
retained
to be re-used for the next measurement.
In summary, in this scheme as is shown in Figure~\ref{flowchart} we
first measure $\sigma_z$ weakly, followed by 
$\sigma_x$
which is again measured weakly and last we make
a projective
measurement of $\sigma_y$. The entire simulation is run on 
identically prepared copies (ensemble size) 
of the state of interest (pure or mixed). The simulation is
repeated many times to avoid statistical errors.

\par A general single qubit state is given by
\begin{eqnarray}
\rho&=&\begin{pmatrix}
\rho_{00}&& \rho_{01}\\
\rho_{10}&& \rho_{11}
\end{pmatrix}\nonumber\\
&=&\rho_{00}\ket{0}\bra{0}+
\rho_{01}\ket{0}\bra{1}+\rho_{10}\ket{1}\bra{0}\nonumber\\
&& +\rho_{11}\ket{1}\bra{1}
\end{eqnarray}
The diagonal elements are known as populations as
they give the probabilities with which the states $\ket{0}$
and $\ket{1}$ are present in the mixture. The off-diagonal
elements are known as coherences as these contain the
phase information of the states $\ket{0}$ and $\ket{1}$.
When the state is coupled to a measurement device,
as discussed above,
the resultant state
after unitary evolution for a strength $\epsilon$, is
\begin{eqnarray}
&&\rho'=\left(\frac{\epsilon}{2\pi}\right)^\frac{1}{2} 
\nonumber \\
&&\left[\int_{-\infty}^\infty dq
\int_{-\infty}^\infty
dq'\rho_{00}\,e^{-\frac{\epsilon\left(q-1\right)^2}{4}}
\,e^{-\frac{\epsilon\left(q'-1\right)^2}{4}}\ket{0}\bra{0}+ 
\right. \nonumber\\
&&
\int_{-\infty}^\infty dq \int_{-\infty}^\infty
dq'\rho_{01}\,e^{-\frac{\epsilon\left(q-1\right)^2}{4}}
\,e^{-\frac{\epsilon\left(q'+1\right)^2}{4}}\ket{0}\bra{1}+
\nonumber\\
&&\int_{-\infty}^\infty dq \int_{-\infty}^\infty
dq'\rho_{10}\,e^{-\frac{\epsilon\left(q+1\right)^2}{4}}
\,e^{-\frac{\epsilon\left(q'-1\right)^2}{4}}\ket{1}\bra{0}+
\nonumber\\
&& \left.
\int_{-\infty}^\infty dq \int_{-\infty}^\infty
dq'\rho_{11}\,e^{-\frac{\epsilon\left(q+1\right)^2}{4}}
\,e^{-\frac{\epsilon\left(q'+1\right)^2}{4}}\ket{1}\bra{1}\right]  
\nonumber\\
&&\otimes \ket{q}\bra{q'}
\label{grand_2}
\end{eqnarray}
Let us consider taking out a member of the ensemble of
system states and then coupling it with the apparatus.
Now when the observer
notes down the meter reading he or she can see a particular
reading which depends upon the initial states of the system
and the meter and the coupling between the two. Though this
process is not well understood and von Neumann's model is
silent about this final step of collapse, it can be thought
of as the action of the projector $\ket{q}\bra{q}$ on the
meter state resulting in the meter reading $q$.

\par The probability density of obtaining the value $q$ for the meter
is therefore given by
\begin{equation}
 P(q)=Tr\left(\ket{q}\bra{q}\rho_{MD}\right)
\end{equation}
where the reduced density operator for the apparatus or the
measuring device (MD)
is obtained by taking a partial trace of the state
$\rho^\prime$ over the system. 
\begin{equation}
 \rho_{MD}=Tr_{system}\left(\rho'\right)
\end{equation}
This probability density can now be used to calculate the
probabilities of possible outcomes. For example,
$P(\sigma_z=1)$ can be
obtained by integrating the probability density from $+a$ to $\infty$.
Thus, the
probabilities with which we obtain $+1$ , $-1$ or ambiguous
readings while measuring in the $z$-basis are
calculated by integrating the above probability densities
from $+a$ to
$\infty$, $-\infty$ to $-a$ and $-a$ to $+a$, respectively
and are given by
\begin{widetext}
\begin{eqnarray}
P(\ket{0})&=&\frac{1}{4}\left[\left(1+z\right)
Erfc{\frac{\left(-1+a\right)\sqrt{\epsilon_1}}{\sqrt{2}}}-\left(-1+z\right)
Erfc{\frac{\left(1+a\right)\sqrt{\epsilon_1}}{\sqrt{2}}}\right]
\nonumber\\
P(\ket{1})&=&\frac{1}{4}\left[-\left(-1+z\right)
Erfc{\frac{\left(-1+a\right)\sqrt{\epsilon_1}}{\sqrt{2}}}+\left(1+z\right)
Erfc{\frac{(1+a)\sqrt{\epsilon_1}}{\sqrt{2}}}\right]
\nonumber\\
P(discard_z)&=&\frac{1}{2}\left[Erf{\frac{\left(-1+a\right)
\sqrt{\epsilon_1}}{\sqrt{2}}}+Erf{\frac{\left(1+a\right)
\sqrt{\epsilon_1}}{\sqrt{2}}}\right]
\end{eqnarray}
\end{widetext}
Further for the second weak measurement, the input
state is the output from the first measurement described by
an ensemble $\rho_1^\prime$. This ensemble is obtained from
the state $\rho^\prime$ given in Equation~(\ref{grand_2}) by
taking a trace over the measuring device (apparatus)
\begin{equation}
\rho^\prime_1= {\rm Tr}_{MD}(\rho^\prime)
\end{equation}

The probabilities with which we obtain the value $+1$, 
$-1$  or
ambiguous readings while measuring in the $\sigma_x$-basis are 
given by, 
\begin{widetext}
 \begin{eqnarray}
P(\ket{\sigma_x;+})&=&\frac{1}{4}e^{-\frac{\epsilon_1}{2}}
\left[\left(-Erf{\left(-1+a\right)
\sqrt{\frac{\epsilon_2}{2}}}+
Erf{\left(1+a\right)\sqrt{\frac{\epsilon_2}{2}}}\right)x+
e^{\frac{\epsilon_1}{2}}\left(
Erfc{\left(-1+a\right)\sqrt{\frac{\epsilon_2}{2}}}+
Erfc{\left(1+a\right)
\sqrt{\frac{\epsilon_2}{2}}}\right)\right]\nonumber\\
P(\ket{\sigma_x;-})&=&\frac{1}{4}e^{-\frac{\epsilon_1}{2}}
\left[\left(Erf{\left(-1+a\right)
\sqrt{\frac{\epsilon_2}{2}}}-
Erf{\left(1+a\right)\sqrt{\frac{\epsilon_2}{2}}}\right)x+
e^{\frac{\epsilon_1}{2}}
\left(Erfc{\left(-1+a\right)
\sqrt{\frac{\epsilon_2}{2}}}+Erfc{\left(1+a\right)
\sqrt{\frac{\epsilon_2}{2}}}\right)\right]\nonumber\\
P(discard_x)&=&\frac{1}{2}
\left[Erf{\frac{\left(-1+a\right)\sqrt{\epsilon_1}}
{\sqrt{2}}}+Erf{\frac{\left(1+a\right)
\sqrt{\epsilon_1}}{\sqrt{2}}}\right]
\end{eqnarray}
\end{widetext}
After this measurement if we trace over the second apparatus
we obtain the ensemble represented through a density
operator $\rho^{\prime\prime}_2$. Lastly we perform a
regular strong (projective) measurement of $\sigma_y$ and
the  probabilities are given by,
\begin{eqnarray}
P(\ket{\sigma_y;+})&=&\frac{1}{2}
\left[1+e^{-\frac{1}{2}\left(\epsilon_1+\epsilon_2\right)}y\right]
\nonumber\\
P(\ket{\sigma_y;-})&=&\frac{1}{2}\left[1-e^{-\frac{1}{2}
\left(\epsilon_1+\epsilon_2\right)}y\right]
\end{eqnarray}
In the above equations, we have used
\begin{eqnarray}
Erf(x)&=&\frac{2}{\sqrt{\pi}}\int_0 ^x e^{-t^2} dt
\nonumber\\
Erfc(x)&=&1-Erf(x)
\end{eqnarray}

These measurements when repeated over the entire ensemble
give us an estimate of the expectation values of
$\sigma_x$, $\sigma_y$ and $\sigma_z$, which in turn help us
locate the 
co-ordinates $(x,y,z)$ of the point inside the
Bloch sphere:
\begin{eqnarray}
z &=& Tr\left(\rho \sigma_z\right)
\nonumber\\
x &=& Tr\left(\rho'_1 \sigma_x\right)e^{\frac{\epsilon_1}{2}}
\nonumber\\
y &=& Tr\left(\rho''_2 \sigma_y\right)
e^{\frac{1}{2}\left(\epsilon_1+\epsilon_2\right)}
\label{estimate}
\end{eqnarray}
where $\rho$, $\rho'_1$ and $\rho''_2$ denote the initial
state of the system and those after the first and second
measurements respectively. We note that $\epsilon_1$ and
$\epsilon_2$ appear in Equation~(\ref{estimate}) 
because we are interested in the expectation
values of $\sigma_x$, $\sigma_y$ and $\sigma_z$ for the
original state $\rho$ of the system. These results are valid
only for small values of $\epsilon_1$ and $\epsilon_2$.  In
subsequent studies we work with the simplification
$\epsilon_1=\epsilon_2 =\epsilon$.

For a scheme based purely on projective measurements, the
ensemble is divided into three equal parts and direct
measurements of $\sigma_x$, $\sigma_y$ and $\sigma_z$ are
performed independently. This leads to a direct estimate of
the expectation values of these operators giving the values
of  $(x,y,z)$ and hence an estimate of the state. The error
in these estimates depends upon the size of the ensemble.
We simulate both these schemes and compare the performance
of our method with the one based on projective measurements.

\subsection{Two Random examples}
To begin with  we perform the simulations on two randomly
generated states  $\rho_1$ and $\rho_2$  given by
\begin{equation}
\rho_1=\frac{1}{2}
\begin{pmatrix}1.399 && \!\!\!\!\!\!\!\!\!-0.385+0.042 i \\
-0.385-0.042 i && 0.601
\end{pmatrix}
\end{equation}
and
\begin{equation}
\rho_2=\frac{1}{2}\begin{pmatrix}1.055 &&\!\!\!\! -0.601-0.398 i \\
-0.601+0.398 i && 0.945
\end{pmatrix}
\end{equation}
On the Bloch sphere these states correspond to $(x=
-0.385;\,
y= -0.042;\, z= 0.397)$ and $(x= -0.601;\, y= 0.398;\, z= 0.055)$
respectively.  Both these states are mixed states with
distance from the origin of the Bloch sphere for $\rho_1$
being $0.555$ and that for $\rho_2$ being $0.723$. Clearly
$\rho_2$ is less mixed than $\rho_1$.  We would like to stress
that these states are randomly chosen.

\begin{figure}[h]
\includegraphics{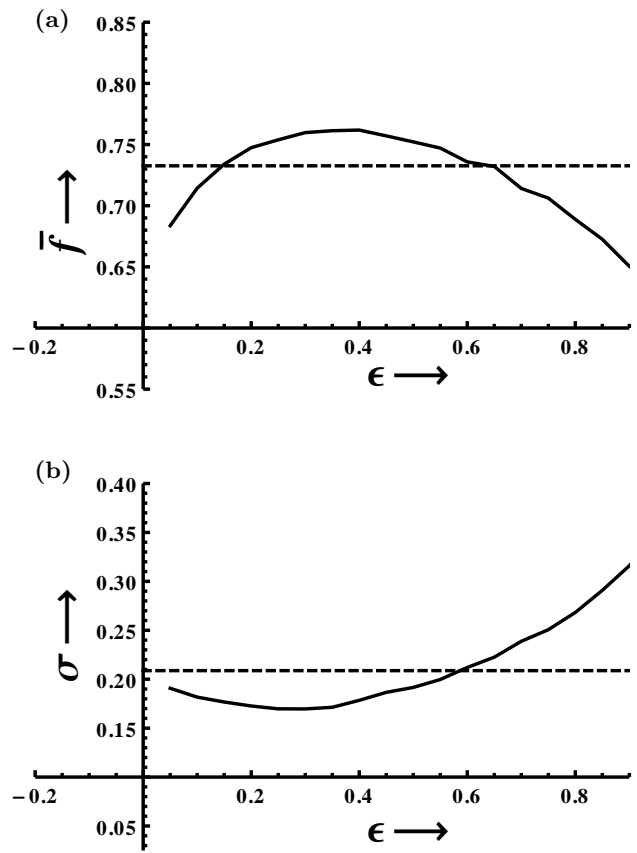}
\caption{Fidelity, $\bar{f}$, ((a)) and standard
deviation $\sigma$ in fidelity ((b)), plotted as a function of
coupling strength $\epsilon$ for a randomly chosen state
$\rho_1$. The size of the ensemble here is  $30$. Weak
measurement (solid line) outperforms
projective measurement (broken black line) for small ensemble
sizes. No values are discarded in this simulation, hence
the discard parameter $a=0$. The straight horizontal dotted
line represents the projective measurement and our scheme
clearly outperforms the projective measurements.}
\label{plot1}
\end{figure}

\begin{figure}[h]
\includegraphics{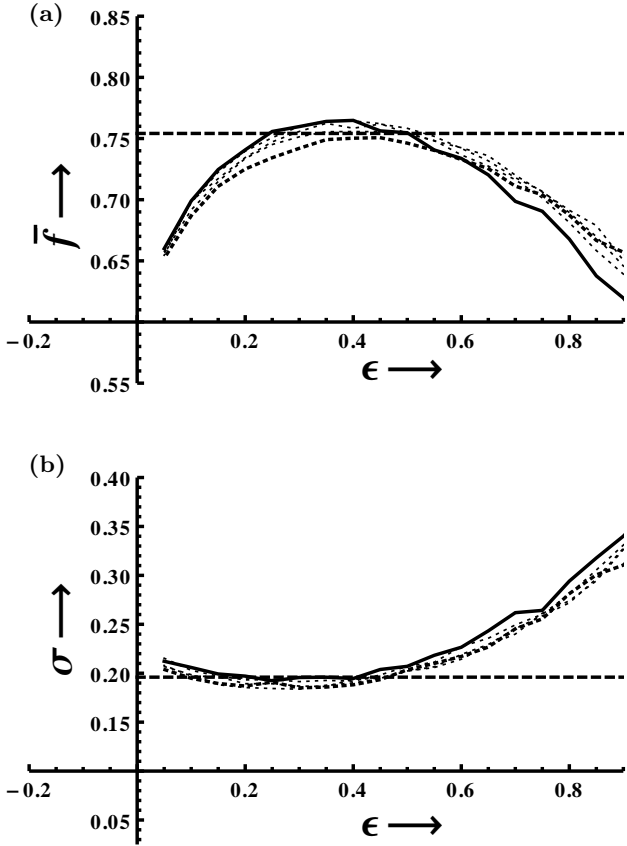}
\caption{Fidelities, $\bar{f}$, ((a)) and standard
deviations $\sigma$ in fidelity ((b)) plotted as a function of
coupling strength $\epsilon$ for a randomly chosen state
$\rho_2$ with an ensemble size of $30$. In each graph different
lines represent different values of
the discard parameter $a$. The discard parameter values plotted are 
$a=0$ (dotted thick line), $a=0.2$ (dotted light line), $a=0.4$ 
(dotted light line),
$a=0.6$ (dotted light line) and $a=0.8$ (solid line).  The solid line
corresponds to the best case where our scheme outperforms the
projective measurements which are represented by the
straight horizontal broken line. The standard deviation graph
for $a=0.8$ (the best case) represented by the solid line 
indicates that the noise in the
tomography based on our scheme is not more than that of
projective measurements.}
\label{plot2}
\end{figure}

Taking ensembles of size 30 and putting
$\epsilon_1=\epsilon_2=\epsilon$ we perform simulations
(using Wolfram Mathematica 9) over 10000 runs and calculate
the individual fidelity for each run. We use the definition
of fidelity defined in Equation~(\ref{fidelity}). The mean
fidelity $\bar{f}$ and the standard-deviation $\sigma$ in
fidelity are then plotted as a function of  $\epsilon$ and in
each case a comparison is made with the same parameters for
the projective measurement case (See Figures~\ref{plot1},
\ref{plot2}). We also vary the breadth of the region in
which  we discard the pointer readings to get an idea of how
it affects the quality of state estimation.

In each case we see that there is an interesting dependence
of the fidelity of the estimate on $\epsilon$ and the
discard parameter $a$. For the case $\rho_1$ the weak
scheme outperforms the projective measurement scheme even
without any discard parameter. On the other hand for
$\rho_2$ we have to increase the discard parameter
considerably to outperform the projective measurements.
However, for another randomly chosen state the scheme may
not outperform the projective measurement scheme.

\par The analysis of the mean fidelity $\bar{f}$ vs $\epsilon$
plots shows that there are states such as $\rho_1$ for which
tomography by weak measurements is more effective than
projective measurements, for small ensemble sizes
(Figure~\ref{plot1}). We note that only in a certain range of
$\epsilon$ values this is true. The reason is not difficult
to see. If $\epsilon$ is large then the state of the system
is destroyed in the very first measurement of $\sigma_z$ and
the subsequent measurements become meaningless. Again, if
$\epsilon$ is made too small, then the overlap of the
``Gaussians'' is too large and a large number of the meter
readings fall in the overlapping region. These meter
readings then cannot be utilized for any useful purpose
under our scheme. The plot of the standard deviation in
fidelity $\sigma$ vs. $\epsilon$ shows that for the optimal
values of $\epsilon$, even standard deviation of fidelity by
weak measurements is less than that for the projective measurements.

\par There are some states, though, for which this method of
estimation does not do better than that by projective measurements.
In some of these cases, the estimation can be improved by
discarding a certain range of meter readings, as was
discussed earlier.  The state $\rho_2$ is an example
of such a case where by increasing the discard parameter we
can outperform the projective measurements. The results for
this state are given in Figure~\ref{plot2}. The
average fidelity becomes better than that for the projective
measurements and the corresponding standard
deviation of fidelity ($\sigma$ plotted as a function of
$\epsilon$) is of the same order as that for projective measurements.
\subsection{Average performance over Bloch sphere}
Encouraged by these results we now move on to test our
scheme on a large number of randomly generated states of a
qubit and look for the average performance of the scheme
over the Bloch sphere.  The  process is carried out for 2000
states generated randomly. We also study the dependence on
ensemble size and use ensemble sizes of $30$, $60$ and $90$.
For each case the simulation is repeated $1000$ times to 
average over statistical fluctuations.

\begin{figure}
\includegraphics{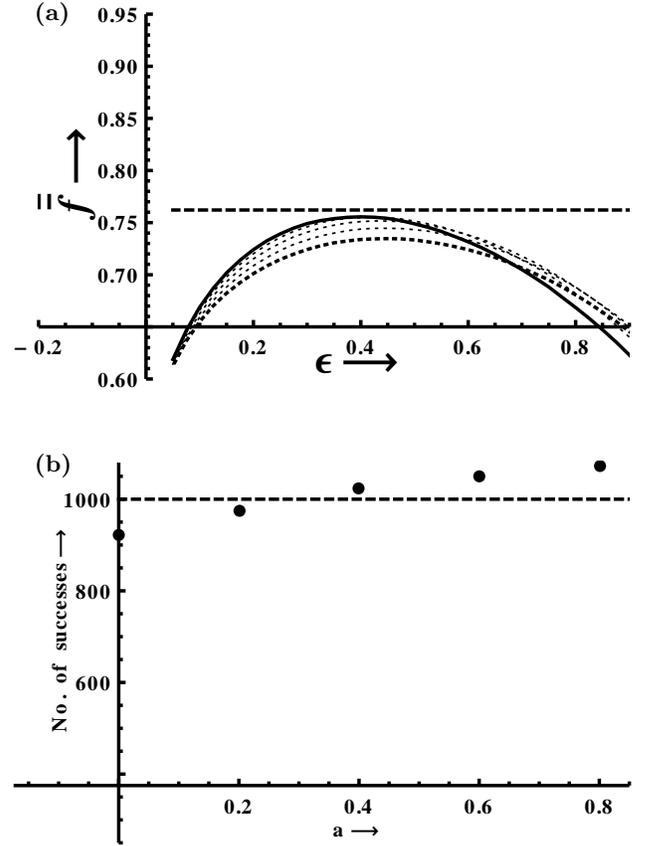}
\caption{(a) Plot of  the mean fidelity $\bar{\bar{f}}$  for
a state with ensemble size $30$ and mean calculated over
1000 runs, further averaged over 2000 randomly chosen
states, as a function of the coupling strength $\epsilon$.
Different curves represent different values of the discard
parameter $a$.  The discard parameter used are  $a=0$
(dotted thick line), $a=0.2$ (dotted line), $a=0.4$ (dotted
line) $a=0.6$
(dotted line) and $a=0.8$ (solid line). The straight dotted line
represents projective measurements. The solid line
comes very close to the projective measurements. (b) Plot of
the number of times our schemes outperform the projective
measurement based scheme for the 2000 randomly chosen states
of the qubit as a function of the discard parameter $a$. The
dotted horizontal line represents the 50\% mark. The
performance of our schemes is better than the projective
measurement schemes when the discard parameter crosses a
certain value (approximately $0.3$).}  
\label{fid_score_30}
\end{figure}

\begin{figure}
\includegraphics{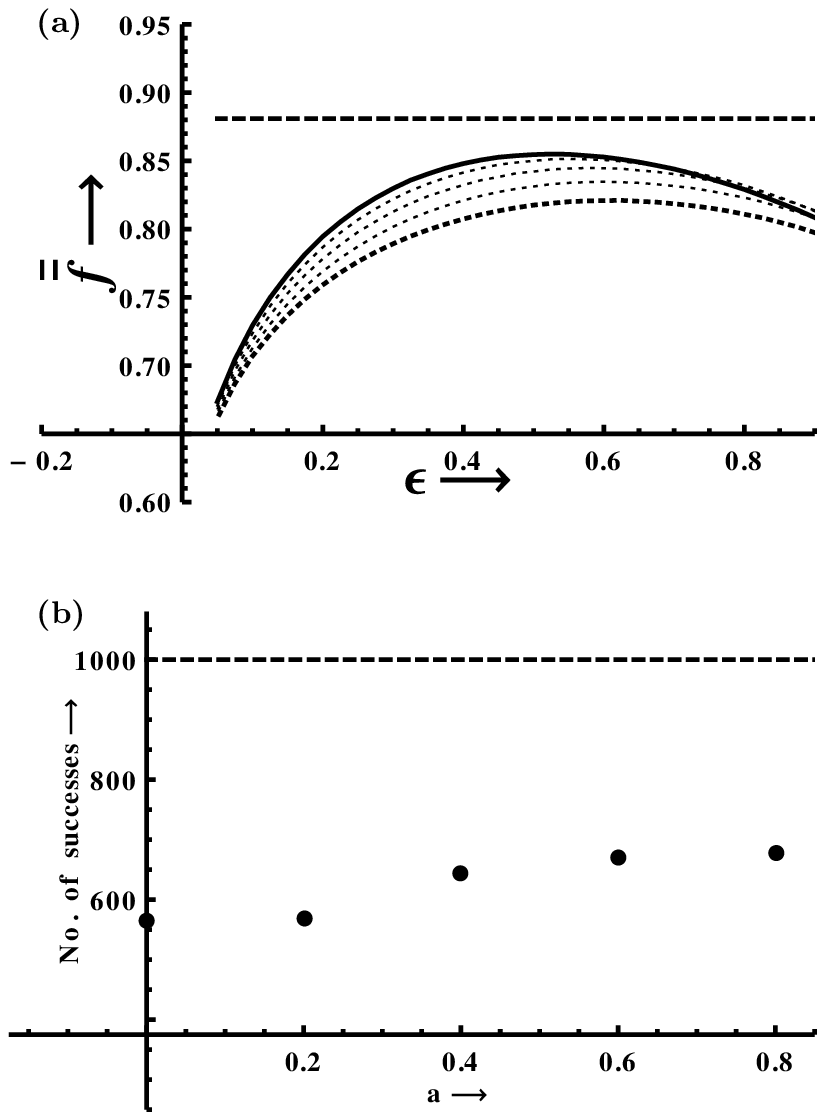}
\caption{(a) Plot of  the mean fidelity $\bar{\bar{f}}$  for
a state with ensemble size $60$ and mean calculated over
1000 runs, further averaged over 2000 randomly chosen
states, as a function of the coupling strength $\epsilon$.
Different curves represent different values of the discard
parameter $a$.  The discard parameter used are  $a=0$
(dotted thick line), $a=0.2$ (dotted line), $a=0.4$ (dotted
line) $a=0.6$
(dotted line) and $a=0.8$ (solid line). The straight dotted line
represents the  projective measurements. 
(b) Plot of
the number of times our schemes outperform the projective
measurement based scheme for the 2000 randomly chosen states
of the qubit as a function of the discard parameter $a$. 
The success rate goes down with an increase in ensemble
size from $30$ to $60$.
}
\label{fid_score_60}
\end{figure}
\begin{figure}
\includegraphics{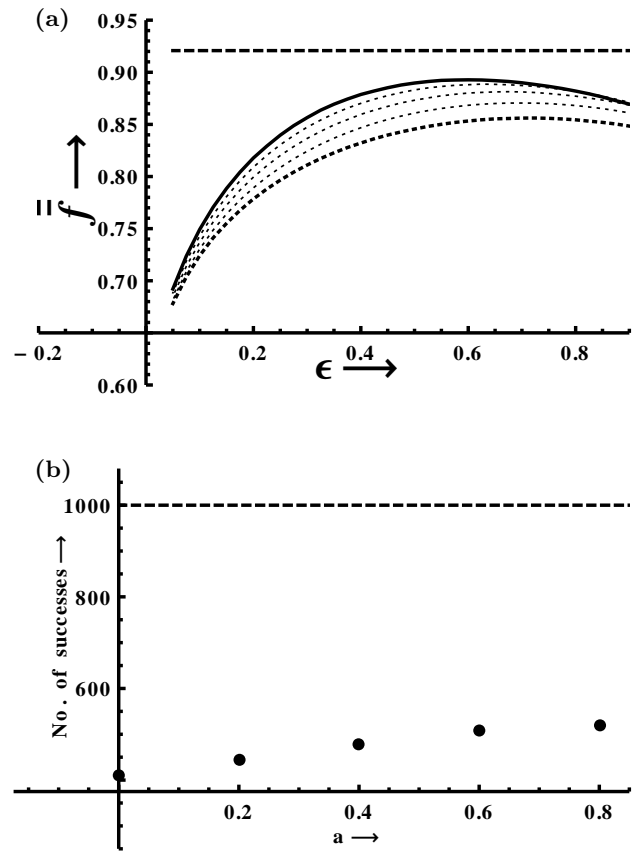}
\caption{(a) Plot of  the mean fidelity $\bar{\bar{f}}$  for
a state with ensemble size $90$ and mean calculated over
1000 runs, further averaged over 2000 randomly chosen
states, as a function of the coupling strength $\epsilon$.
Different curves represent different values of the discard
parameter $a$.  The discard parameter used are  $a=0$
(dotted thick line), $a=0.2$ (dotted line), $a=0.4$ (dotted
line) $a=0.6$
(dotted line) and $a=0.8$ (solid line). The straight dotted line
represents the  projective measurements. 
(b) Plot of
the number of times our schemes outperform the projective
measurement based scheme for the 2000 randomly chosen states
of the qubit as a function of the discard parameter $a$. 
The success rate further decreases with an increase in ensemble
size to 90.
}
\label{fid_score_90}
\end{figure}
While we average the fidelity over all states to obtain the
average fidelity we also keep track whether the scheme
outperformed or underperformed as compared to the
projective measurement scheme in each case.  For the ensemble size of
$30$, the results of this simulation are presented in two
different ways in Figure~\ref{fid_score_30}. We calculate
the mean fidelities averaged over these states,
$\bar{\bar{f}}$, with and without discard, which are then
plotted against $\epsilon$ in Figure~\ref{fid_score_30}(a).
This graph shows an improvement as we increase the amount of
discard. We also present our results through a
score plot, where we compute the number of states out of 2000
starting states for which our scheme outperforms the
projective measurement scheme.  The score plot is described
in Figure~\ref{fid_score_30}(b).  Interestingly, this number
crosses the 50\% mark for a threshold value of the discard
parameter.

When a study of mean fidelity, averaged over 2000
states, $\bar{\bar{f}}$ vs $\epsilon$ was done, it turns out
that although on the average the performance of projective
measurements is better, if ambiguous meter readings are
discarded, then the number of states for which our
tomography scheme is successful, goes up. In fact, number of
successes out of 2000 for the discard parameter values of
0, 0.2, 0.4, 0.6 and 0.8 
are 923, 973, 1023, 1051 and 1071,
respectively.
This we think is a clear evidence that our
scheme has the potential of unearthing more information than
projective measurements under certain circumstances.
In particular,
if we are given $30$ copies of a unknown state of
a qubit, our scheme will be a better choice for carrying out
state tomography.

We now turn to testing our scheme with increasing ensemble size.
We repeat the simulation in exactly the same way for the
cases of ensemble size $60$ and $90$.
The results are presented in a similar way in
Figures~\ref{fid_score_60} and~\ref{fid_score_90}.
Increasing the ensemble size clearly reduces the
efficacy of our scheme as compared to projective measurements.
The score plots show that our scheme outperforms the
projective measurement scheme for ensemble sizes of $60$ and
$90$ for lesser number of states and the number is less than
$50\%$. Therefore we conclude that our scheme is preferable
only when we have a small ensemble size. We would
like to clarify that this not due to statistical
fluctuations as we have taken the average over a large number of
runs even when the ensemble size is small.
\subsection{States with $\langle\sigma_y \rangle=0$}
\begin{figure}[h]
\includegraphics{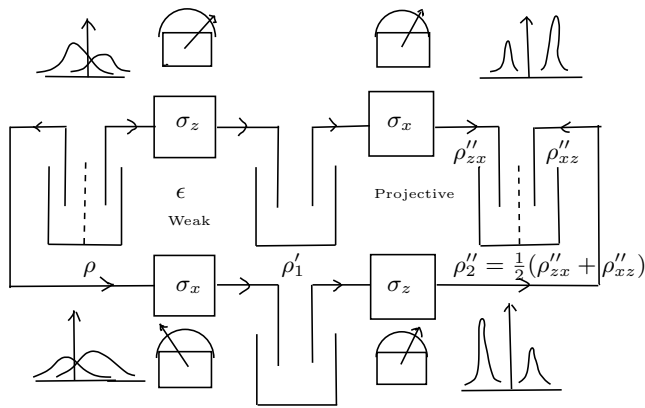}
\caption{
Block diagram of the weak
measurement based state estimation scheme
applied to the case when the states lie on a disk in the
Bloch sphere with $\langle \sigma_y\rangle=0$. For such
states only two measurements are performed, namely, a weak
measurement followed by a projective measurement. 
To achieve symmetry we divide the ensemble into
two halves and for the first half we carry out weak
measurement of $\sigma_z$ followed by a projective
measurement of $\sigma_x$ and for the second half we reverse
the order of the measurements. For the first half the final
density operator after both the measurements is
$\rho_{zx}^{\prime\prime}$ and the same for the second half
is $\rho_{xz}^{\prime\prime}$. The density  operator for
the entire ensemble after the measurement thus is
$\rho^{\prime\prime}_2=
\frac{1}{2}(\rho^{\prime\prime}_{zx}+
\rho^{\prime\prime}_{xz})$}
\label{flowchartplane} 
\end{figure}
\begin{figure}[h]
\includegraphics{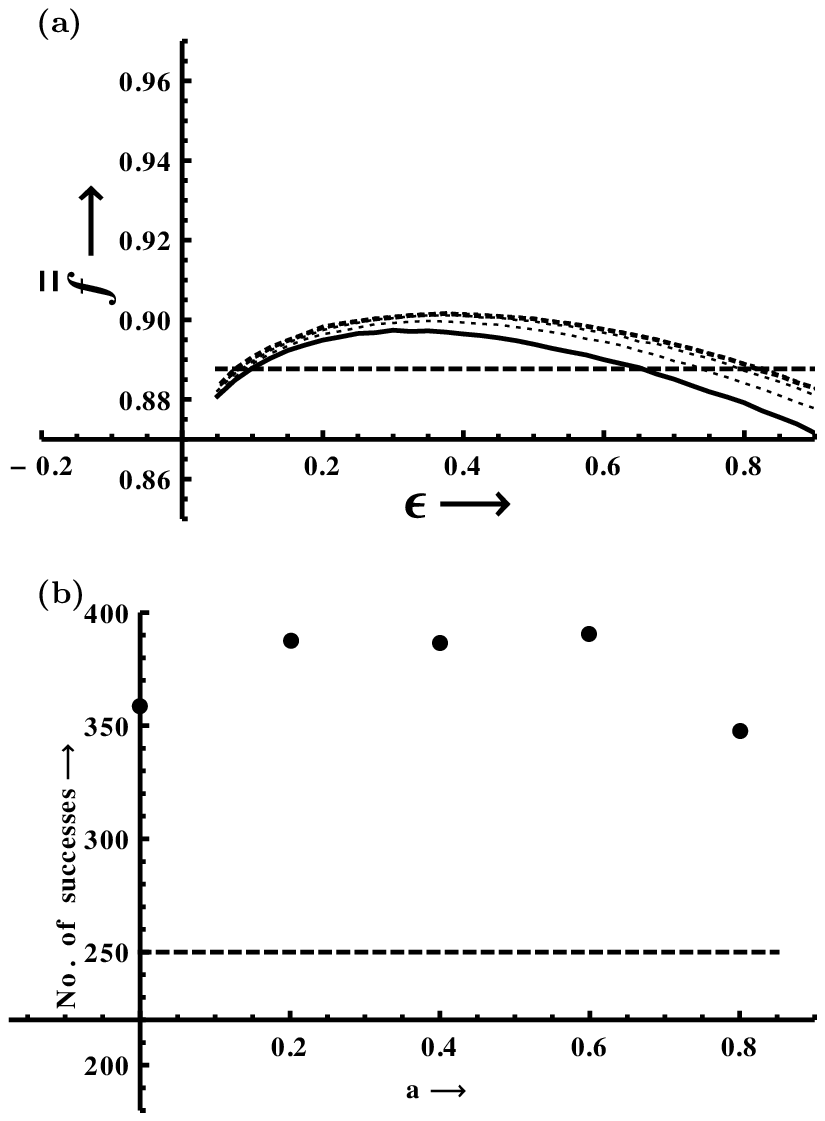}

\caption{Results of state estimation for 500 randomly
generated states on the disk with $\langle \sigma_y\rangle
=0$ for an ensemble size of $30$ with each state averaged over
1000 runs. The average fidelity as a
function of $\epsilon$ for discard parameter values $0$,
$0.2$, $0.4$, $0.6$ and $0.8$ are shown in part (a). In part
(b) the score plot is displayed where we plot the number of
successes out of 500 as function of discard parameter $a$.}
\label{pl_30_fid}
\end{figure}
\begin{figure}[t]
\includegraphics{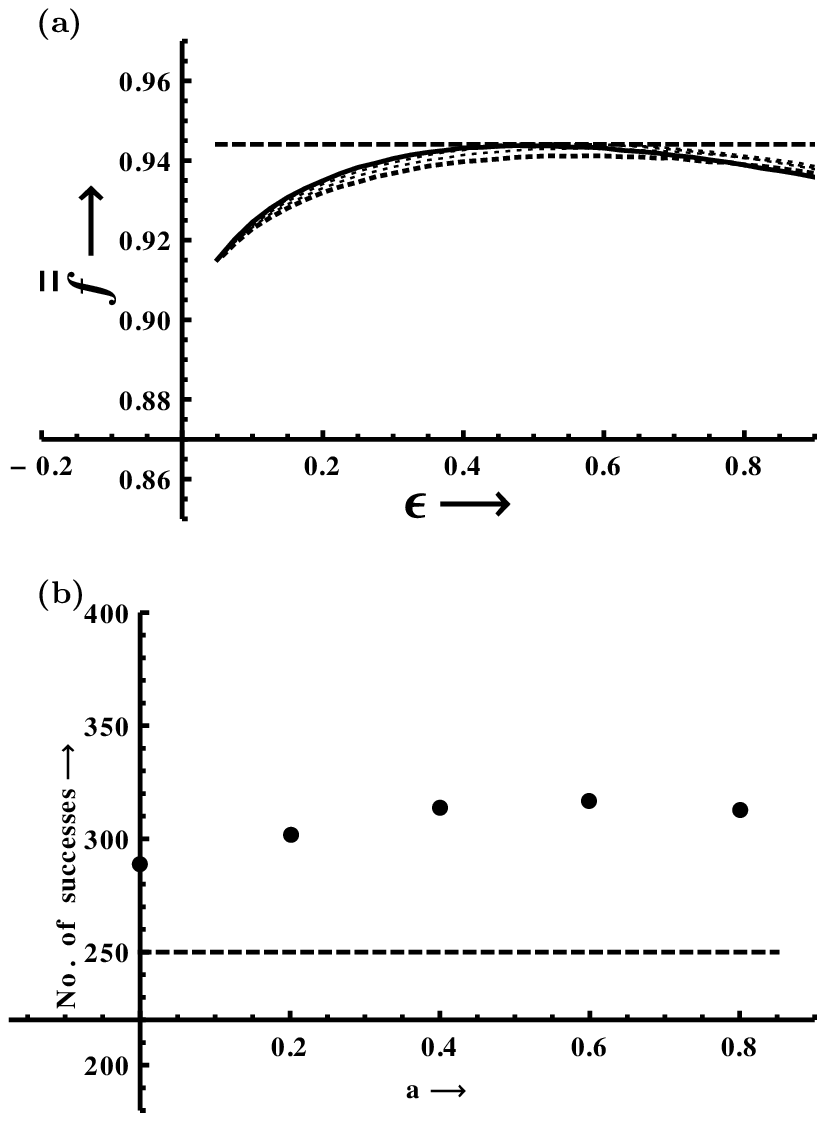}
\caption{Results of state estimation for 500 randomly
generated states on the disk with $\langle \sigma_y\rangle
=0$ for an ensemble size of $60$ with each state averaged over
1000 runs. The average fidelity as a
function of $\epsilon$ for discard parameter values $0$,
$0.2$, $0.4$, $0.6$ and $0.8$ are shown in part (a). In part
(b) the score plot is displayed where we plot the number of
successes out of 500 as function of discard parameter $a$.}
\label{pl_60_fid}
\end{figure}
\begin{figure}[t]
\includegraphics{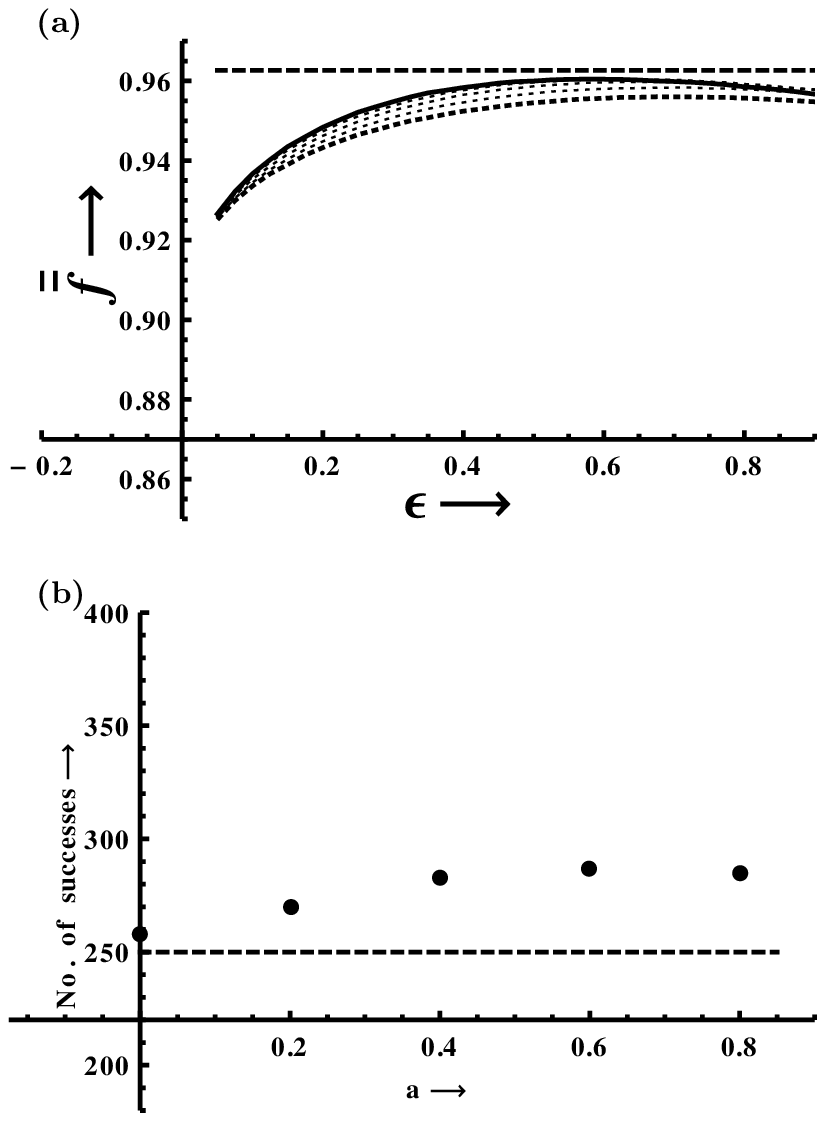}
\caption{Results of state estimation for 500 randomly
generated states on the disk with $\langle \sigma_y\rangle
=0$ for an ensemble size of $90$ with each state averaged over
1000 runs. The average fidelity as a
function of $\epsilon$ for discard parameter values $0$,
$0.2$, $0.4$, $0.6$ and $0.8$ are shown in part (a). In part
(b) the score plot is displayed where we plot the number of
successes out of 500 as function of discard parameter $a$.}
\label{pl_90_fid}
\end{figure}

We now turn to a subset of states in the Bloch sphere,
namely the states with $\langle \sigma_y \rangle=0$. These
states form a disk perpendicular to the $\hat{y}$ axis and
passing through the origin. The set contains pure states
which lie on a circle and mixed states all the way to
maximally mixed state which lie in the interior of the disk.
Once it is known that a state belongs to this set, estimations
of only two parameters are required to know the state. This
can be achieved by measuring $\langle \sigma_x\rangle$ and 
$\langle \sigma_z\rangle$.
Given a finite size  ensemble of identically
prepared states belonging to this set, how do we estimate the
state and how well can we do it? If we employ projective
measurements we can divide the ensemble into two equal parts
and measure $\langle \sigma_x \rangle$ on one half of the
system and measure $\langle \sigma_z \rangle$ on the other
half. In our weak measurement based scheme with state
recycling, we again divide the ensemble into two equal parts.
On the first half we carry out a weak measurement of
$\sigma_x$ of strength $\epsilon$ followed by a projective
measurement of $\sigma_z$ while on the second half we
reverse the order where we carry out a weak measurement of 
$\sigma_z$ followed by a projective measurement of
$\sigma_x$. 
The flowchart of this measurement scheme is shown in
Figure~\ref{flowchartplane}.

While simulating the scheme, we choose ensemble sizes of
$30$, $60$ and $90$ and compare the state estimation
efficacy of our scheme with the projective measurement
scheme. For each ensemble size, we generate $500$ random
states in the set,  and repeat the estimation $1000$ times for
each state.  The results for ensemble size $30$ are
displayed in Figure~\ref{pl_30_fid}. The results are
presented exactly in the same way as we did in the previous
section. The results of ensemble sizes $60$ and $90$ are
presented in Figures~\ref{pl_60_fid} and~\ref{pl_90_fid}
respectively.

For this subset of states the weak measurement based scheme
does much better. The score plots show that the scheme
outperforms the projective measurements in all the three
cases. The relative efficacy reduces as the ensemble size is
is increased.  
\section{Concluding Remarks}
\label{conc}
In this work we have used weak measurements to carry out
quantum state tomography on finite-sized (pure or mixed)
one-qubit ensembles. We have shown that in such schemes,
recycling of states is possible, where one
makes more than one measurement on a single copy before discarding
a given member of the ensemble of identically prepared
states. In general when coupling strengths are small, the
pointer positions may overlap, making the outcome of the
measurement ambiguous.  We have introduced a discard
parameter such that the outcomes with most ambiguity
are discarded. We have carried out an optimization of the
scheme to improve its efficacy with respect to 
the coupling strength $\epsilon$ and the discard parameter
$a$.

Over a randomly chosen subset of qubit states, our scheme
performs better than the scheme based on projective
measurements. We demonstrate this by showing that the weak
measurement based scheme works better for more than 50\%
of the randomly chosen cases for small ensemble sizes.
For a subset of states on the Bloch sphere where we take a
disk with $\langle \sigma_y\rangle =0$ the scheme does very
well and is almost always preferable over projective
measurements. As the ensemble size increases the relative
efficacy of our scheme decreases as seen in the comparative
results for varying ensemble sizes.

It is true that an experimenter will not know a priori
whether, for a given unknown state, which scheme out of
projective measurement and weak measurement will be more
suitable.  However, the experimenter will be able to make an informed
choice, depending on knowledge about the ensemble size. In
the particular case where the $\sigma_y $ polarization is
zero or for that matter any particular polarization is known
to be zero, our method will be a better choice for state
estimation.

This has opened up an interesting possibility
of estimating quantum states and extracting information from
quantum ensembles using weak measurements.  We would also
like to mention that the original context in which the weak
measurements were introduced was related to weak value and
post selection.  However, we do not carry out any post
selection and do not use the weak value. We only use the
weak nature of the measurement to recycle the states.

In many physical situations, the apparatus is weakly
coupled with the system and  hence our scheme may find a
natural application for such measurements. In another
direction, a natural extension of this scheme on two qubits,
where entangled states are possible, can lead to interesting
possibilities. In particular, one may be able to detect
quantum entanglement by such schemes.  A more detailed
discussion of this and related results will be taken up
elsewhere.
%
\end{document}